\documentclass[onecolumn,draftcls]{IEEEtran}

\ifCLASSINFOpdf
  \usepackage[pdftex]{graphicx}
  \usepackage{epstopdf}
  \graphicspath{{./pic/}}
  \DeclareGraphicsExtensions{.pdf,.jpeg,.png}
\else
  \usepackage[dvips]{graphicx}
  \DeclareGraphicsExtensions{.eps}
\fi

\ifCLASSOPTIONcompsoc
    \usepackage[tight,normalsize,sf,SF]{subfigure}
\else
    \usepackage[tight,footnotesize]{subfigure}
\fi

\usepackage{amssymb}
\usepackage[cmex10]{amsmath}
\usepackage{algorithm}
\usepackage{algorithmicx}
\usepackage{algpseudocode}
\usepackage{amsmath}
\usepackage{array}
\usepackage{mdwmath}
\usepackage{mdwtab}
\usepackage{amsfonts}
\usepackage{url}
\usepackage{color}

\begin{document}
\title{SDPMN: Privacy Preserving MapReduce Network Using SDN}

\author{\IEEEauthorblockN{He Li, Hai Jin}\\
\IEEEauthorblockA{Services Computing Technology and System Lab \\ Cluster and Grid Computing Lab\\
School of Computer Science and Technology\\
Huazhong University of Science and Technology, Wuhan, 430074, China\\
\{heli, hjin\}@hust.edu.cn}}

\maketitle

\begin{abstract}
%Mobile cloud computing provides flexibility and economically efficient service through mobile terminals, including laptops, tablets and smartphones. To manage mobile cloud computing, using software defined networking (SDN) is scalable and convenient. Unlike ordinary wired networking, the rule placement problem in the mobile cloud networking is much more complex with moving devices. In this paper, we modeling this problem named rule placement in a SDN enabled mobile cloud network. We also propose a heuristic algorithm to solve this problem. From the evaluation based on simulation, our algorithm performs good performance than the existing placement method.
MapReduce is a popular programming model and an associated implementation for parallel processing big data in the distributed environment. Since large scaled MapReduce data centers usually provide services to many users, it is an essential problem to preserve the privacy between different applications in the same network. In this paper, we propose SDPMN, a framework that using \textit{software defined network} (SDN) to distinguish the network between each application, which is a manageable and scalable method. We design this framework based on the existing SDN structure and Hadoop networks. Since the rule space of each SDN device is limited, we also propose the rule placement optimization for this framework to maximize the hardware supported isolated application networks. We state this problem in a general MapReduce network and design a heuristic algorithm to find the solution. From the simulation based evaluation, with our algorithm, the given network can support more privacy preserving application networks with SDN switches.\\
\end{abstract}

\begin{IEEEkeywords}
Software Defined Network, MapReduce, Privacy
\end{IEEEkeywords}

%\IEEEpeerreviewmaketitle

\section{Introduction}
%With more and more convenient network accessing, people often using mobile terminals to access service, including laptops, tablets and smartphones. Mobile cloud computing is a flexibility and economically efficient way to provide services to these types of device. Since network management in cloud computing is important to quality of service, companies begin to adopting SDN technology to support more and more complex networks. SDN technology separates network control as an independent plan with a centralized controller and programmable network devices. The network communication in SDN can be full controlled by the centralized controller and 

MapReduce is the most popular programming model for large data processing with parallel and distributed environment \cite{Dean2008}. Since the large clusters usually provide services to different users all over the Internet, it is hard to preserve the privacy of user's data with traditional network structure \cite{Zhou2010}.

There are some existing works focus on the privacy problem in MapReduce \cite{McSherry2009}\cite{Roy2010}. These works usually bring some additional overhead to influence the performance of MapReduce. Adopting \textit{software defined network} (SDN) or the OpenFlow protocol is another way to preserve privacy between users in the MapReduce network, which can distinguish network links between each users \cite{Mogul2010}.

Privacy leakage might take place in the Layer 2 switches, including the virtual switch in cloud environment, by the sniffer and ARP poisoning-based attack \cite{Nam2010}. SDN based network virtualization is an efficient method to isolated network to resist the potential data leakage in the data forwarding path. While virtualized the whole network is complex and resource-intensive, it is much appropriate that isolating the network forwarding paths in the specific network level. 

SDN is an approach of computer networking, which allows the network administrator to manage network services through abstraction of hardware level functionality \cite{McKeown2009}. All hardware devices are managed by a centralized controller, and the network applications are deployed on this controller to implement different network functions. In the privacy preserving, the network is divided to distinguished virtual networks for each user \cite{Chowdhury2009}. 

In this paper, we propose a framework named \textit{Software Defined Privacy Preserving MapReduce Network} (SDPMN), in which the entire network are divided to distinguished groups and each groups are isolated in the Layer 2 switches by SDN technology. In this framework, all packets are forwarded by SDN switches directly with different rules. We design this framework with existing SDN technology and Hadoop structure. 

SDN enabled switches usually equip \textit{ternary content addressable memory} (TCAM) for forwarding network packets \cite{McKeown2008}. Since it is power hungry, expensive and takes up quite a bit of silicon space, the storage space of TCAM in each device is limited. Ordinarily, as it is hard to place all rules in the switches, many rules are stored in the controller with the performance drawback \cite{Qazi2013}.

In SDPMN, since forwarding paths are implemented by rules in the related SDN switches directly, the space for storing these rules is not enough with existing SDN hardware. Therefore, with the design of the SDPMN framework, we state rule placement optimization problem to maximize the application number with limited rule space. After that, we propose a heuristic algorithm to solve this problem. To evaluate the efficiency of our algorithm, we take some simulation and analyze the result to compare the existing method. 

 The main contributions of this paper are summarized as follows.
\begin{itemize}
\item We propose a privacy preserving scenario in the leaf switches to forward application packets by the SDN forwarding rules directly instead of the traditional ARP protocols.

\item We model the privacy preserving MapReduce network scenario and state the problem to find an optimized rule placement that maximum application groups for preserve user privacy are stored in the switches with minimum performance loss.
\item We design a heuristic algorithm and find the solution for the placement problem. 
\end{itemize}

The rest of this paper is summarized as follows. We discuss the motivation of the work in Section \ref{sec:motiv}. In section \ref{sec:statement}, we introduce the framework design and the rule placement problem statement. We also introduce our heuristic algorithm to solve the placement problem in this section. In section \ref{sec:eva}, we evaluate the demonstration of our design to verify our methodology. In the last section, we conclude our work and discuss the future work.

\section{Related Works}
In this section, we introduce some existing works focusing on the isolation and security of MapReduce network. First, we discuss the works on the network virtualization. Then, we introduce some works on the privacy preserving in MapReduce networks. 

Network virtualization is a technology to isolate the network to different virtual networks for different users or applications.

Bavier et al.  \cite{Bavier2006} propose the virtual network infrastructure, which allows network researchers to evaluate their protocols and services in a realistic environment with a high degree network control. From their work, the virtualized network performs good efficiency to support experiments and other applications. This work only focuses on the IP level without any consideration on the Level 2 switches.

ShadowNet  \cite{Chen2009} is another network evaluation infrastructure based on the network virtualization. Similarly, it also uses the router virtualization and designs a network controller to manage the virtual networks. Meanwhile, the network virtualization can provide safe environment for users. However, they did not describe the details how to ensure testing safely.

Webb et al.  \cite{Webb2011} introduces a network that individual application can control the topology switching for deciding best path to route data among their nodes. In their paper, it is mentioned that a $k$ isolation problem to isolate networks by limited hardware capacity. They use an allocation to find a topology that connects the task's nodes that is minimally shared with other tasks.

Drutskoy et al.  \cite{Drutskoy2013} design an architecture named FlowN, which can give each application or task the illusion of its own address space, topology, and controller. Actually, the FlowN architecture is a network virtualization without the SDN context. Even it mentioned that they improve the network scalability, they did not describe the security and rule placement for many applications.

The network virtualization is very important to provide isolated virtual network for different application. The problem is, they focus on how to create virtual networks, which needs substantial resource like the space of TCAM or others.

Privacy is a major topic of concern whenever large amounts of sensitive data are used in MapReduce. Even though there are many privacy preserving works to protect the data privacy, a few of works focus on the security issue of MapReduce. 

Airavat  \cite{Roy2010} is privacy preserving system that provides strong security and privacy guarantees for distributed computations. By integration of mandatory access control and differential privacy, Airavat makes data providers control their security policy to protect their privacy. Since it is an efficient work to resist the leakage of data privacy, there is no consideration about the potential leakage in the network.

Wei et al.  \cite{Wei2009} proposed a secure scheme named SecureMR which focuses at protecting the computation integration issue of MapReduce. SecureMR detects malicious behavior of mappers by sending same tasks to multiple mappers, and compares the result consistency. This work needs the master node and reduce be secure, it seems a little hard if some malicious users exist in the MapReduce cluster.

Huang et al.  \cite{huang2012} proposes a trusted services by verifying the MapReduce result. They focus on detecting cheating services under the MapReduce environment based on watermark injection and random sampling methods. While the privacy leakage is hard to influence the computation result, we do not consider result verification an efficient method to preserve privacy.

\section{Motivation}
\label{sec:motiv}
In this section, we first describe the threat model to introduce the privacy preserving problem in the multiple user MapReduce networks. Then, with this threat model, we discuss the privacy preserving MapReduce network.

\subsection{Threat Model}

\begin{figure}[h]
\centering
\includegraphics[width=2.5in]{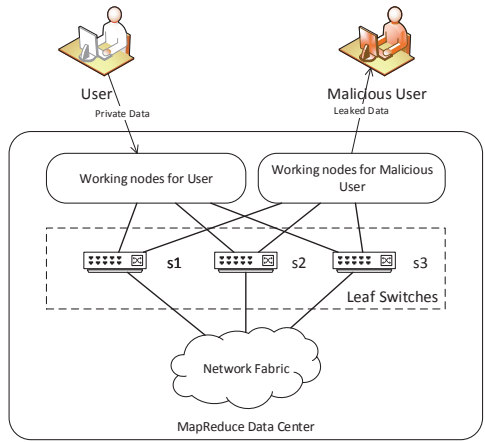}
\caption{Example of the potential threat in MapReduce network}
\label{fig:thret_mode}
\end{figure}

For better understanding the privacy problem in MapReduce networks, we describe an example to introduce the threat model. This example shows that a MapReduce network will leak data privacy between different users. As shown in Fig. \ref{fig:thret_mode}, there are two users in a MapReduce network: one has some privacy data and the other one is a potential malicious user. Both of them submit their applications to this MapReduce data center. In the traditional network, the MapReduce nodes, including Map nodes and Reduce nodes, are connected by many Layer 2 access switches (leaf switches) \cite{Al-Fares2008}. These access switches are connected by some interconnection switches or more complex network.

First, the MapReduce system such as Hadoop assigns some nodes as working nodes for the normal user and the malicious user. After assignment, the normal user sends the private data to the working nodes. After that, these data are translated to the switches ($s1$, $s2$ and $s3$) in MapReduce procedures. However, it is possible that the working nodes of the malicious user also connect to some of these switches ($s1$, $s2$ and $s3$). Since the leaf switches are usually Layer 2 switches, which are connected by Ethernet or other cheap solutions, the working nodes for malicious user, who uses sniffer or other attacks like Man-in-the-middle attack, can get the packets transferred in these switches. Even though the main implementations of MapReduce like Hadoop adopt secure connections between working nodes, it is possible to get the privacy data by some existing attack methods \cite{Khan2012}. 

\subsection{Privacy Preserving with SDN}
\begin{figure}[h]
\centering
\includegraphics[width=2.5in]{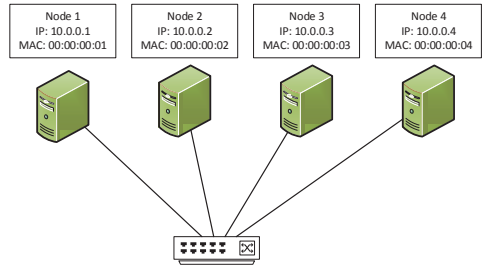}
\caption{Example to illustrate forwarding rules for privacy preserving}
\label{fig:privacy_flow}
\end{figure}

\begin{table}[h]
\center
\caption{Ports in Fig. \ref{fig:privacy_flow}}
\begin{tabular}{|l|l|}
\hline
Port & Node                              \\ \hline
Port 1 & Node 1           \\ \hline
Port 2   &  Node 2             \\ \hline
Port 3 & Node 3 \\ \hline
Port 4 & Node 4 \\ \hline
\end{tabular}
\label{tab:node_port}
\end{table}
\begin{table}[h]
\center
\caption{Forwarding Rules in Fig. \ref{fig:privacy_flow}}
\begin{tabular}{|l|}
\hline
Forwarding Rules \\ \hline
IF IP = 10.0.0.1 THEN MAC $\leftarrow$ 00:00:00:01, PORT $\leftarrow$ 1                             \\ \hline
IF IP = 10.0.0.2 THEN MAC $\leftarrow$ 00:00:00:02, PORT $\leftarrow$ 2          \\ \hline
IF IP = 10.0.0.3 THEN MAC $\leftarrow$ 00:00:00:03, PORT $\leftarrow$ 3            \\ \hline
IF IP = 10.0.0.4 THEN MAC $\leftarrow$ 00:00:00:04, PORT $\leftarrow$ 4 \\ \hline

\end{tabular}
\label{tab:flows}
\end{table}
SDN is a network structure that all switches are controlled by the controller. As a result, if we deploy SDN switches as all switches in the network, the path of each packet can be defined by the applications rather than the basic network protocols. In the MapReduce network, with SDN network, we can isolate the network by the forwarding rules \cite{Shin2013}. 

To study the SDN network for privacy preserving in MapReduce network, we use an example to illustrate isolation by forwarding rules as shown in Fig. \ref{fig:privacy_flow}. There are node 1, 2, 3 and 4, connect to an SDN switch. The connected port of each node is described in the table shown in TABLE \ref{tab:node_port}. To preserve the privacy, we use four rules to limit the potential sniffer or other attacker in the node beside the destination. With the rules illustrated in the table shown in TABLE \ref{tab:flows}, each packet is only forwarded to the port connected to the destination IP address. As an example, each packet sent to node 2 with the destination IP address 10.0.0.2 will be forwarded to port 2 after setting the correct MAC address. 

Unlike the traditional network virtualization, since the potential threats are not relevant to the upper network level such as IP packet forwarding, we only use the SDN rules to isolate network in the Layer 2 switches. Considering the nodes connected to each leaf switch are fixed with the number of available ports, it is acceptable to place the rules for network isolation between applications.
 
\section{Framework Design and Problem statement}
\label{sec:statement}
\begin{figure}[h]
\centering
\includegraphics[width=2.5in]{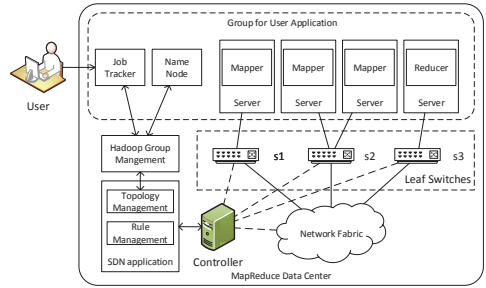}
\caption{SDPMN framework and the MapReduce network}
\label{fig:system_design}
\end{figure}

In this section, we first introduce the design of SDPMN framework. With this framework, we state this problem to optimize the maximum groups with limited rule space in each switch.

\subsection{Framework Design}
From the main structure shown in Fig. \ref{fig:system_design}, the framework consists of two parts: group management and the SDN application.

For each application, group management gets the information about all mapper, reducer and other related nodes in the network. After that, group management divides these nodes to groups. The nodes for one application are organized to one group. When the user submits an application to the Job Tracker, Job Tracker informs group management the application information. The name node also connects group management to inform the information about mappers and reducers. 

With the information of applications and their mappers and reducers, group management maintains a table to store the correspondences between applications and servers. When the nodes of application are changed by the Job Track, the new node information will be transferred to group management to update the group table dynamically. 

The SDN application generates rules to isolate the network in Layer 2 switches with the group table from group management and the network topology. In this SDN application, one module is used for the management of network topology and the other modules generate rules and manages these rules. 

Since the network topology will be changed frequently in some virtual machine based networks, the topology management will update the relationship between switches, nodes and groups dynamically. When the topology is updated, rule management generates the newest rules. Since the switch space is limited to store all needed rules, the SDN application selects parts of rules for placement to switches. For the groups whose rules are not placed in the switches, it will bring additional overhead to forward the packets in these groups to guarantee the privacy. Since it is relevant to the controller implementation and the network design  \cite{Yu2010}\cite{Moshref2012}, we do not discuss this overhead in the placement. At last, the SDN application updates the subset of rules to the corresponding switches. We introduce the rule placement problem in the final part of this section.

\begin{figure}[h]
\centering
\includegraphics[width=2.5in]{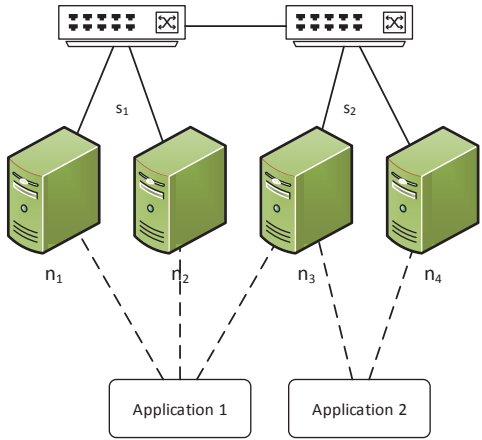}
\caption{Example of the SDPM}
\label{fig:statement}
\end{figure}

\subsection{Problem Statement}

\begin{table}[h]
\center
\caption{Notations in the rule placement problem}
\begin{tabular}{ll}

Notation & Description                               \\ \hline
$N_k$             & Node set for each MapReduce application            \\ 
$n_i$             &  Node in set $N_k$              \\ 
$S_k$             & Switch set for each MapReduce application     \\ 
$s_j$             & Switch in set $S_k$\\
$C$           & Capacity of each switch              \\ 
$R$             & Rule space cost for each rule            \\ 
$\mathbb{N}$ & Number of application groups supported by switches \\
\end{tabular}
\label{tab:notation}
\end{table}

Now we discuss the rule placement problem in SDPMN framework. Since existing data centers are not dedicated for MapReduce usage, it is no need to control forwarding of each packet. For each MapReduce application, the nodes for Mapper and Reducer are different. Therefore, for these different node sets, the network should be reconfigured with different rules to define the forwarding path of each packet. We propose a MapReduce driving privacy preserving network shown in Fig. \ref{fig:statement}.

In this network, we first define $N_1$, $N_2$, ..., $N_n$ to denote the node set for each MapReduce application.  For each node set, $N_k$, we use $n_i\in N_k$ to denote each node included in $N_i$. In Fig. \ref{fig:statement}, the node set $N_1$ of Application 1 is $\{n_1, n_2, n_3\}$ and the node set $N_2$ of Application 2 is $\{n_3, n_4\}$. With these node sets, we define $S_1$, $S_2$, ... , $S_n$ to denote the set of leaf switches for each application. For example, in Fig. \ref{fig:statement}, the switch set $S_1$ of Application 1 is $\{s_1, s_2\}$ and the switch set $S_2$ of Application 2 is $\{s_2\}$. For each switch set $S_i$, we also define $s_j\in S_i$ to denote each switch included in $S_i$. We use value $X_{ij}$ to denote the connections between nodes and switches as (\ref{eq:connection}).
\begin{equation}
\label{eq:connection}
X_{ij}=\left\{\begin{matrix}
0 & \ no\ connection\ between\ n_i\ and\ s_j \\ 
1 & \ n_i\ connected\ to\ s_j
\end{matrix}\right.
\end{equation}

For privacy preserving, we consider the number of rules for each node in each application is the same and use $R$ to denote this rule count. Since the hardware space of each SDN switch is usually too expansive to support enough rules, as shown in (\ref{eq:capacity}), we define the number $C$ to denote the capacity of each switch and the capacity of each switch $s_j$ cannot support rule placement of all applications. 
\begin{equation}
\label{eq:capacity}
R\sum_{k=1}^n\sum_{i\in N_k}X_{ij} > C \text{, for } s_j \in S
\end{equation} 

Therefore, in the rule placement, we can only place parts of rules of the switches with the best efficiency. We define set $\mathbb{N}$ to denote the node sets whose rules are placed in switches. With this set, the rule number of each switch is less than its capacity as (\ref{eq:placement}). 

\begin{equation}
\label{eq:placement}
R\sum_{k=1}^{|\mathbb{N}|}\sum_{i\in N_k}X_{ij} \leq C \text{, for } s_j \in S
\end{equation}

From the model of existing works, we consider that each group has the same security priority to simplify the placement problem \cite{Roy2010}. Therefore, with the capacity limitation, we state the rule placement problem in the SDPMN that, with given node sets of each application, find a set $\mathbb{N}$ in which rules of each node are placed in switches and maximize the size of $\mathbb{N}$.    

%\subsection{Hardness analysis}
%\begin{theorem}
%The rule placement problem is NP-hard
%\end{theorem}
%
%We prove the NP-hardness of the rule placement problem by reducing the well-know 0-1 knapsack problem defined as follows.
%
%\textbf{0-1 knapsack problem}: Let there be $n$ objects, $a_1$ to $a_n$ where $a_i$ has a value $p_i$ and weight $w_i$. A vector of binary variables $x_i$ having meaning as follows.
%\[
%x_i = \begin{cases}
%1 & \text{ if object } i \text{ is carried; }  \\ 
%0 & \text{ otherwise. }
%\end{cases}
%\]
%The maximum weight that we can carry in the knapsack is $W$. All binary vectors $x$ satisfying the constraint as follows.
%\[
%\sum_{i=1}^{n}w_i x_i \leq W \text{,}
%\]
%the one which maximizes the sum of value
%\[
%\sum_{i=1}^{n}p_i x_i \text{.}
%\]
%

\subsection{Algorithm Design}
\label{sec:alg}

\begin{algorithm}[t]
\caption{The greedy algorithm for rule placement problem.}
\label{alg:placement}
\begin{algorithmic}[1]
\State Sort node sets $N_1$, $N_2$, ..., $N_n$ that $|S_1| <|S_2| <...<|S_n|$.
\label{sort}
\State $\mathbb{N} \leftarrow \emptyset $
\label{N}
\For {$N_k$ in $N_1$, $N_2$, ..., $N_n$}
\label{begin}
\State $p \leftarrow 0$
\For {$n_i$ in $N_k$}
\State $placed \leftarrow True$
\For {$s_j$ in $S_k$}
\label{place}
\If {$X_{ij} = 1$}
\If {$C_j \geq R$}
\label{constrant}
\State {$C_j \leftarrow C_j - R$}

\Else 
\State $placed \leftarrow False$
\State break
\EndIf
\EndIf
\EndFor
\label{placed}
\If {$placed = False$}
\State Break
\Else
\State $p\leftarrow p+1$
\EndIf
\EndFor
\If {$p = |N_k|$}
\label{put}
\State $\mathbb{N} \leftarrow \mathbb{N} \cup \{N_k\}$
\EndIf
\label{afterput}
\EndFor
\label{end}
\end{algorithmic}
\end{algorithm} 

To solve the placement problem, we propose a heuristic algorithm with greedy strategy shown in Algorithm \ref{alg:placement}. Since each group has the same privacy priority, the main procedure of this algorithm is to place the rules from small group to large group.

First, we sort all node sets by the size of their corresponding switch sets in line \ref{sort}. After sorting, we define the empty set $\mathbb{N}$ in line \ref{N} as the result of the algorithm. From line \ref{begin} to line \ref{end}, we use a \textit{for} loop to place rules of each node set iteratively. In this loop, we select the $N_k$ who has the smallest size of the corresponding switch set in unallocated node sets. For each node $n_i$ in $N_k$, we traverse each switch in the corresponding switch set $S_k$ from line \ref{place} to \ref{placed}. In line \ref{constrant}, we check whether the switch has enough space for rule placement of $n_i$. If there is not enough space for placing rules of $n_i$, it means the rule placement for $N_k$ is failed. With this failure, the algorithm breaks the loop and places rules of the next node set. If success, the capacity of the switch $s_j$ is decreased by $R$. From line \ref{put} to line \ref{afterput}, if the rules of nodes in $N_k$ are all placed, $N_k$ is put to the result set $\mathbb{N}$.

From this algorithm, we can easily get the worst time complexity is $\mathcal{O}(n\sum_{k=1}^{n}|N_k||S_k|)$. For most of the applications, since the number of nodes is much smaller than the scale of the nodes in the entire data center, the complexity can be simplified to $\mathcal{O}(n^2)$.  

\section{Evaluation}
\label{sec:eva}
\begin{figure}
\centering
\begin{minipage}{0.49\textwidth}
	\centering
	\subfigure[Maximum number of supported application groups with light workload ]{
		\includegraphics[width=1\textwidth]{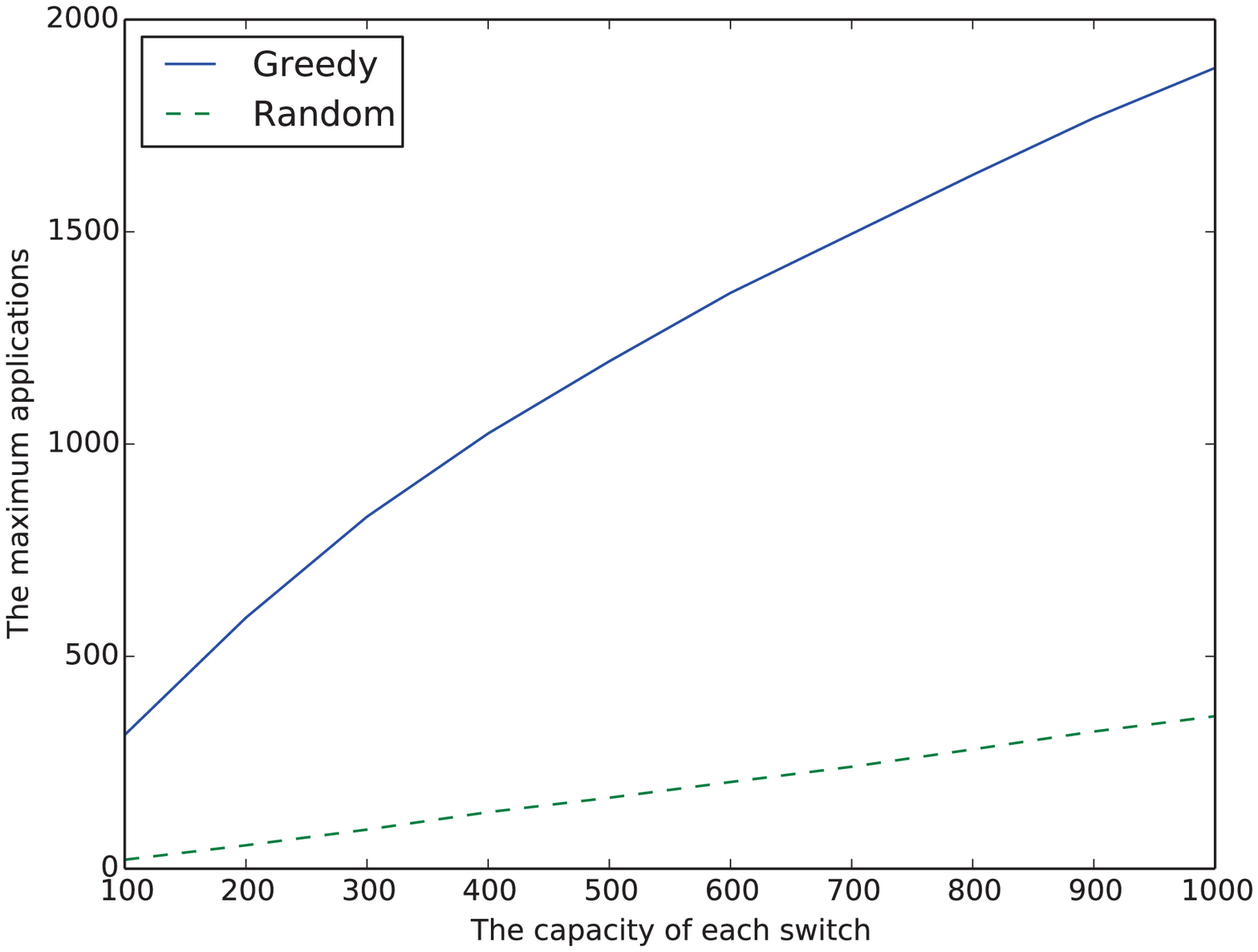}
	\label{fig:result_com1}}
\end{minipage}
\hfill
\begin{minipage}{0.49\textwidth}
	\centering
	\subfigure[Maximum number of supported application groups with heavy workload ]{
		\includegraphics[width=1\textwidth]{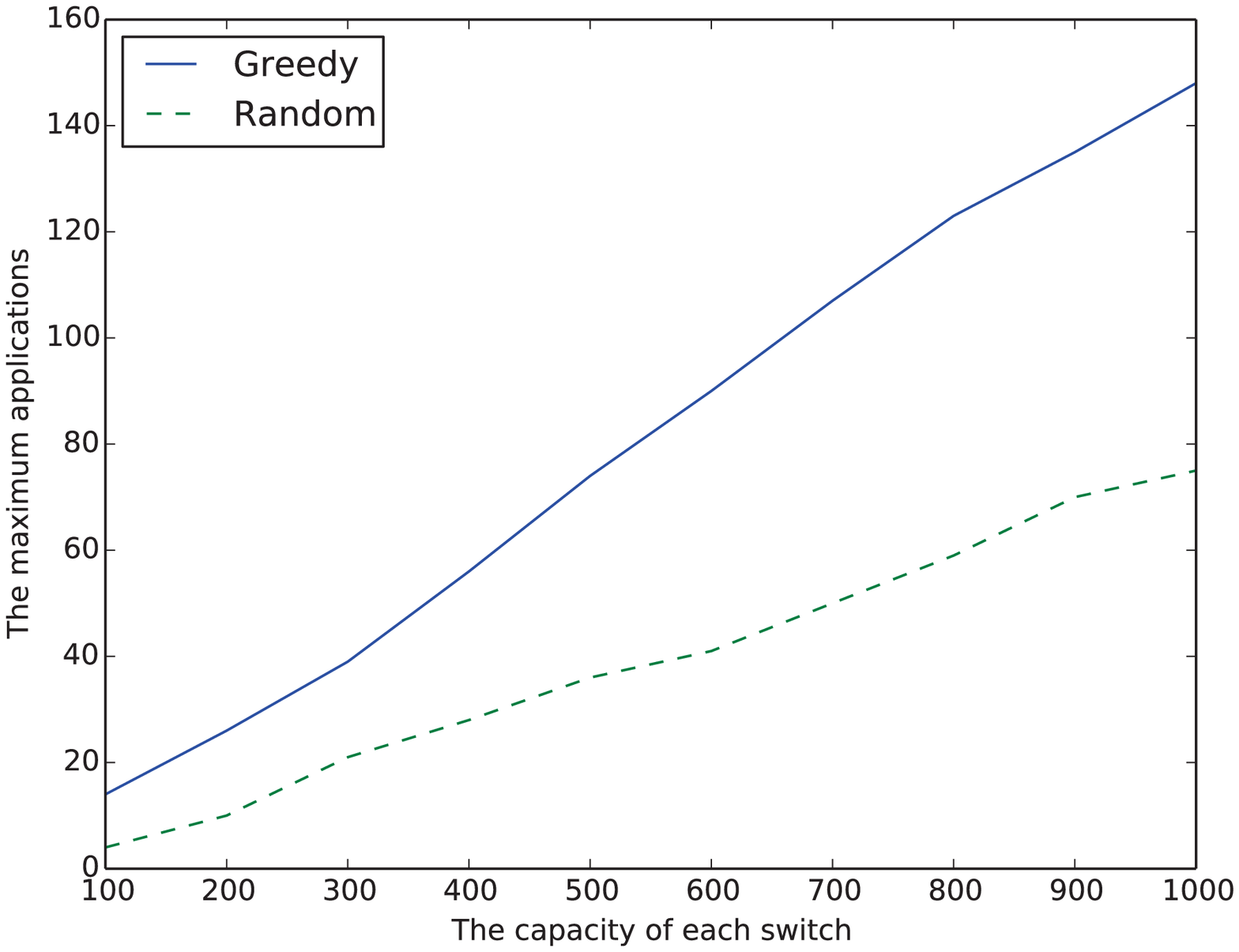}
	\label{fig:result_com2}}
\end{minipage}
\caption{Algorithm performance in the normal data center network}
\label{fig:eva_normal}
\end{figure}

\begin{figure}
\centering
\begin{minipage}{0.49\textwidth}
	\centering
	\subfigure[Maximum number of supported application groups with light workload]{
		\includegraphics[width=1\textwidth]{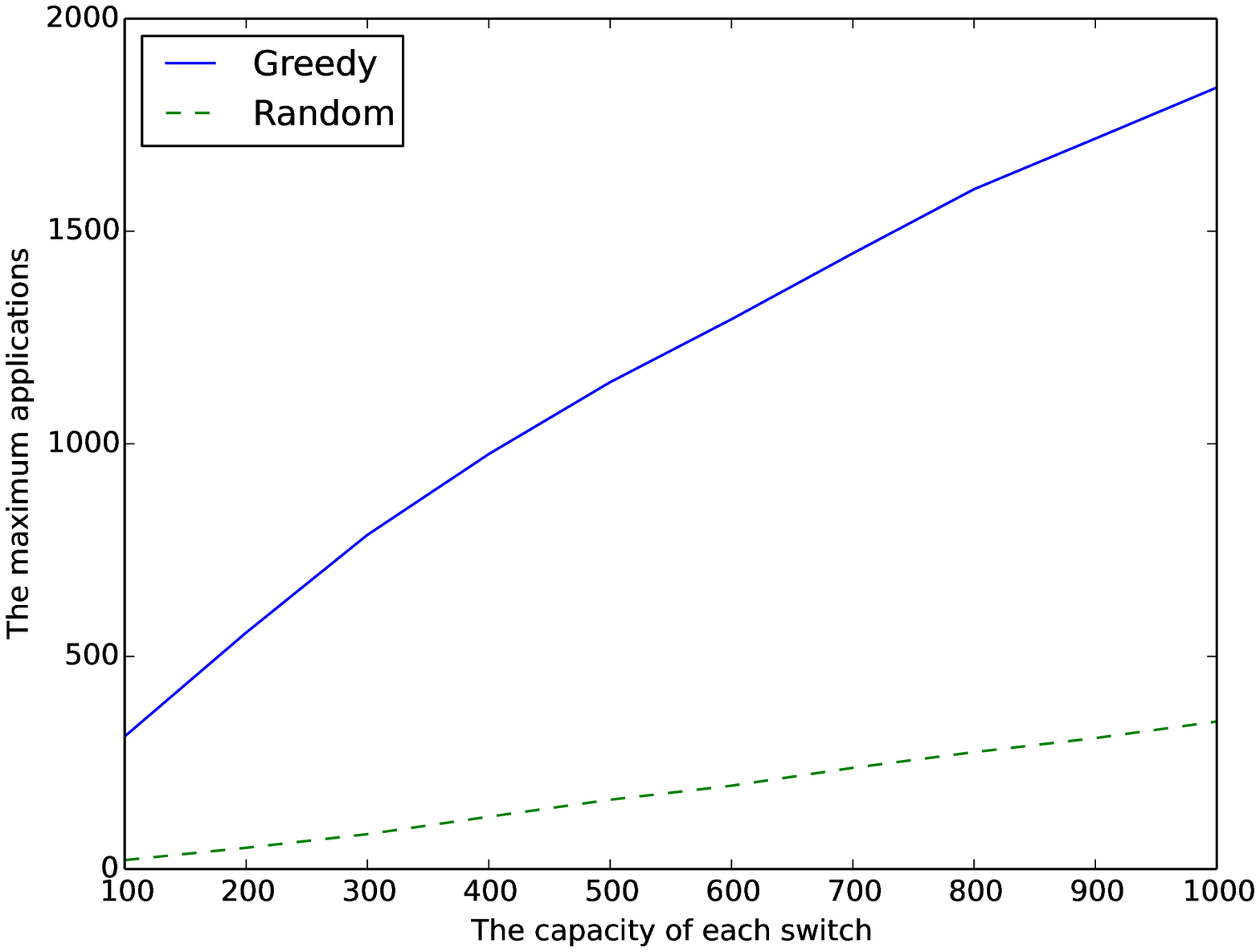}
	\label{fig:result_com3}}
\end{minipage}
\hfill
\begin{minipage}{0.49\textwidth}
	\centering
	\subfigure[Maximum number of supported application groups with heavy workload]{
		\includegraphics[width=1\textwidth]{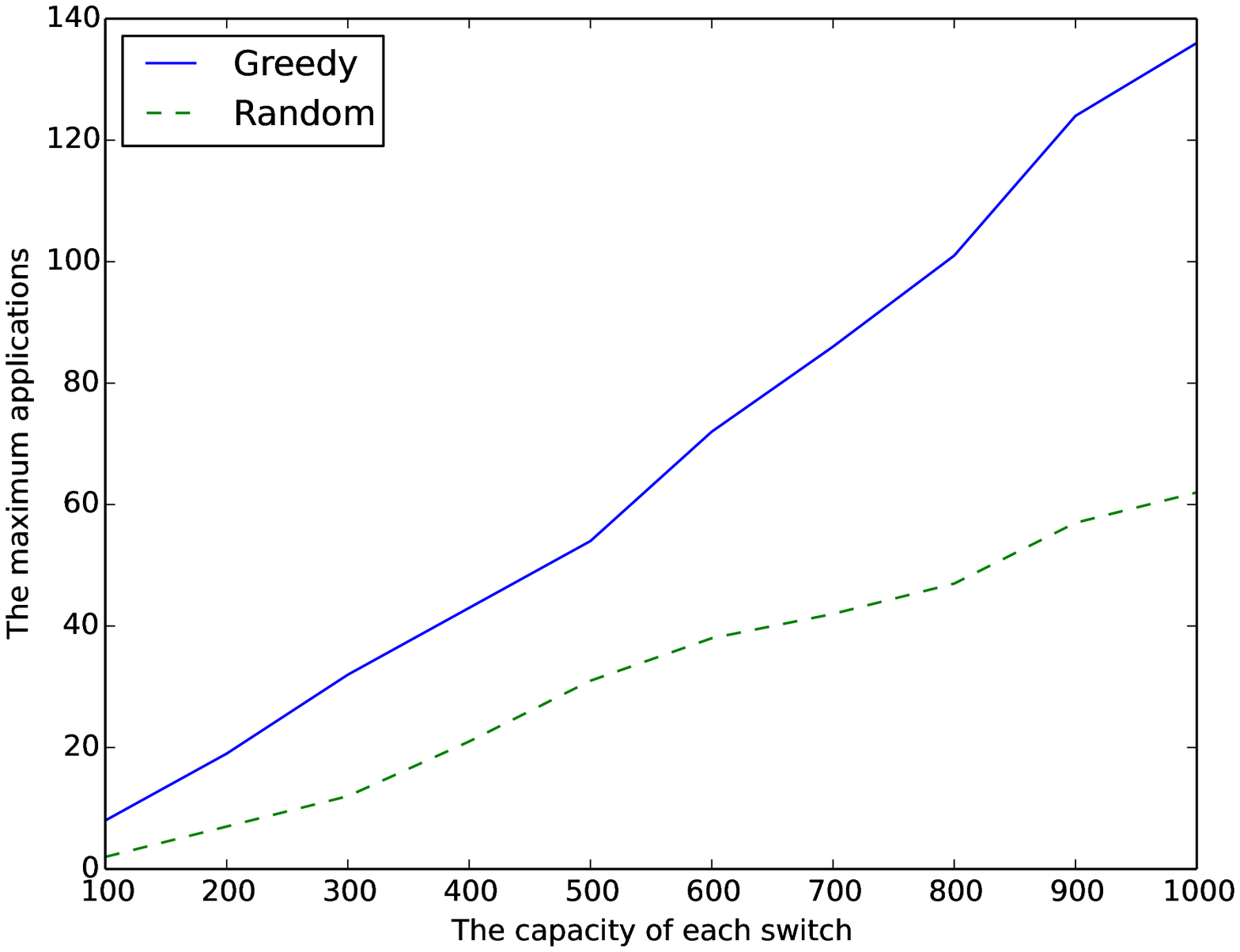}
	\label{fig:result_com4}}
\end{minipage}
\caption{Algorithm performance in the cloud environment network}
\label{fig:eva}
\end{figure}

In this section, we evaluate the performance of the greedy algorithm discussed in section \ref{sec:alg} by simulation. To simulate large scale network topology, we implement some python scripts with networkx library. We defined two different network in the simulation: one network is for simulating the networks in normal MapReduce data center and the other network is for simulating the cloud environment. We also use two types of node scale to simulate the light workload and heavy workload.

First, we introduce the simulation setting of these different networks and node scale. Usually, in the cloud environment, there are more nodes connected to each leaf switch than the normal data center. For the normal network, we set 20 nodes connected to each leaf switch. For the cloud environment, we set 400 nodes connected to each leaf switch. Meanwhile, the number of nodes of the normal network is set to 1000 while the number of nodes in the cloud environment is set to 20000. Both of these networks have the same number of leaf switches which is set to 50. We define two types of node scale for light workload and heavy workload. For the light workload, the number of working nodes is set from 10 to 100 in average distribution. For the heavy workload which needs more working nodes, the number is set from 100 to 200 in the average distribution. To simplify the simulation, we set the cost of each rule to 1 and the maximum space of each switch is set to 1000. This simplification is from the existing SDN switches. Usually, to implement a forwarding rule, it needs about several entries in OpenFlow protocol and the entries supported by normal switch are a few of thousands \cite{Curtis2011}.

With the simulation settings, we evaluate the performance of our algorithm in these networks with different node scale. We first test the performance of the algorithm with the normal data center setting. Since there is no related algorithm for this problem, we use a random placement as the comparison. The performance is shown in Fig. \ref{fig:eva_normal}. With light workload, the greedy placement shows much better than random placement with different rule space shown in Fig. \ref{fig:result_com1}. The maximum number of supported groups is near 2000 with the greedy placement when the rule space of each switch is set to 1000. However, when the workload becomes heavy and needs more working nodes, the supported groups are much less as shown in Fig. \ref{fig:result_com2}. Although the greedy placement performs better than the random placement, the maximum supported groups is less than 160, which is only 8\% of the result with light workload by the same rule space.

With more nodes increased in the cloud data center network condition, the performance of rule placement is almost the same with the normal network condition shown in Fig. \ref{fig:eva}. With light workload setting, the maximum supported number of application groups by the greedy placement are near the 6 times than the number by the random placement shown in Fig. \ref{fig:result_com3}. When the workload increased, the supported groups by the greedy algorithm placement are no more than 140 with 1000 rule space in each switch shown in Fig. \ref{fig:result_com4}. From the result of the simulation, with the light workload of small scale of working nodes in each group, the greedy placement has good efficiency in both normal data center and cloud environment network. For the heavy workload, it seems better to use some task placements rather than simple rule placement optimization. 

Unlike the random placement, in our greedy algorithm, the size of each group is considered. Therefore, more groups with small size can be placed firstly. For the small size groups, since the probability of overlapping between groups is small, the number of supported groups is near to the ideal placement. However, with the average size of groups increased and the frequently overlapping between groups, the number of groups supported by the network is greatly decreased. This is the reason why the groups supported in the network by the greedy placement are only two times than that of random placement.

\section{Conclusion and Future Work}
In this paper, we propose a framework named SDPMN. In this framework, the network forwarding is guaranteed by SDN rules to avoid the potential threats that leak privacy data to malicious users in the same network. Since the rule space of each switch is limited, we state the problem to maximize the number of application groups whose rules are placed in the SDN switches. We propose a heuristic algorithm with the greedy strategy and evaluate this algorithm by simulation. From the simulation results, with the greedy placement algorithm, the given network can support more application groups than random placement. Since the application groups and users are varied in different periods, we will discuss the dynamic placement in the future work. Meanwhile, if the application needs more nodes from the data center network, the rule space of each switch is hard to support many application groups. Considering there are many repeating and redundant rules between application groups, we will take some optimization in the future work.

\section*{Acknowledgment}
This work is supported by National 973 Basic Research Program of China under grant No. 2014CB340600.

%\bibliographystyle{IEEEtran}
%\bibliography{SDMC}

\end{document}